\documentclass[aps,plb,twocolumn,showpacs,superscriptaddress,groupedaddress]{revtex4}

\usepackage{dcolumn}  
\usepackage{bm}       
\usepackage{amssymb}  
\usepackage{comment}  
\usepackage{setspace}  

\usepackage[english]{babel}
\usepackage[dvips]{pstcol}
\usepackage[dvips]{graphicx}
\usepackage{epsfig}
\usepackage{latexsym}
\usepackage{amsmath}
\usepackage{longtable}
\usepackage{multirow}
\usepackage{dcolumn}
\usepackage{rotating}
\usepackage{color}
\usepackage{subfigure}
\usepackage{fancyhdr}

\def\lsim{\mathrel{\rlap{\lower4pt\hbox{\hskip1pt$\sim$}}\raise1pt\hbox{$<$}}}
\def\gsim{\mathrel{\rlap{\lower4pt\hbox{\hskip1pt$\sim$}}\raise1pt\hbox{$>$}}}
\def\MET{{\mbox{$E\kern-0.57em\raise0.19ex\hbox{/}_{T}$}}}
\def\METwithSpace{{\mbox{$E\kern-0.57em\raise0.19ex\hbox{/}_{T}~$}}}

\newcommand{\met}       {\mbox{\ensuremath{\slash\kern-.7emE_{T}}}}

\newcommand{\ttbar}     {\mbox{$t\bar{t}$}}

\newcommand{\comphep}   {\sc{c}\rm{omp}\sc{hep}}

\begin{document}
\hspace{5.2in} \mbox{Fermilab-Pub-11/104-E}

\affiliation{Universidad de Buenos Aires, Buenos Aires, Argentina}
\affiliation{LAFEX, Centro Brasileiro de Pesquisas F{\'\i}sicas, Rio de Janeiro, Brazil}
\affiliation{Universidade do Estado do Rio de Janeiro, Rio de Janeiro, Brazil}
\affiliation{Universidade Federal do ABC, Santo Andr\'e, Brazil}
\affiliation{Instituto de F\'{\i}sica Te\'orica, Universidade Estadual Paulista, S\~ao Paulo, Brazil}
\affiliation{Simon Fraser University, Vancouver, British Columbia, and York University, Toronto, Ontario, Canada}
\affiliation{University of Science and Technology of China, Hefei, People's Republic of China}
\affiliation{Universidad de los Andes, Bogot\'{a}, Colombia}
\affiliation{Charles University, Faculty of Mathematics and Physics, Center for Particle Physics, Prague, Czech Republic}
\affiliation{Czech Technical University in Prague, Prague, Czech Republic}
\affiliation{Center for Particle Physics, Institute of Physics, Academy of Sciences of the Czech Republic, Prague, Czech Republic}
\affiliation{Universidad San Francisco de Quito, Quito, Ecuador}
\affiliation{LPC, Universit\'e Blaise Pascal, CNRS/IN2P3, Clermont, France}
\affiliation{LPSC, Universit\'e Joseph Fourier Grenoble 1, CNRS/IN2P3, Institut National Polytechnique de Grenoble, Grenoble, France}
\affiliation{CPPM, Aix-Marseille Universit\'e, CNRS/IN2P3, Marseille, France}
\affiliation{LAL, Universit\'e Paris-Sud, CNRS/IN2P3, Orsay, France}
\affiliation{LPNHE, Universit\'es Paris VI and VII, CNRS/IN2P3, Paris, France}
\affiliation{CEA, Irfu, SPP, Saclay, France}
\affiliation{IPHC, Universit\'e de Strasbourg, CNRS/IN2P3, Strasbourg, France}
\affiliation{IPNL, Universit\'e Lyon 1, CNRS/IN2P3, Villeurbanne, France and Universit\'e de Lyon, Lyon, France}
\affiliation{III. Physikalisches Institut A, RWTH Aachen University, Aachen, Germany}
\affiliation{Physikalisches Institut, Universit{\"a}t Freiburg, Freiburg, Germany}
\affiliation{II. Physikalisches Institut, Georg-August-Universit{\"a}t G\"ottingen, G\"ottingen, Germany}
\affiliation{Institut f{\"u}r Physik, Universit{\"a}t Mainz, Mainz, Germany}
\affiliation{Ludwig-Maximilians-Universit{\"a}t M{\"u}nchen, M{\"u}nchen, Germany}
\affiliation{Fachbereich Physik, Bergische Universit{\"a}t Wuppertal, Wuppertal, Germany}
\affiliation{Panjab University, Chandigarh, India}
\affiliation{Delhi University, Delhi, India}
\affiliation{Tata Institute of Fundamental Research, Mumbai, India}
\affiliation{University College Dublin, Dublin, Ireland}
\affiliation{Korea Detector Laboratory, Korea University, Seoul, Korea}
\affiliation{CINVESTAV, Mexico City, Mexico}
\affiliation{FOM-Institute NIKHEF and University of Amsterdam/NIKHEF, Amsterdam, The Netherlands}
\affiliation{Radboud University Nijmegen/NIKHEF, Nijmegen, The Netherlands}
\affiliation{Joint Institute for Nuclear Research, Dubna, Russia}
\affiliation{Institute for Theoretical and Experimental Physics, Moscow, Russia}
\affiliation{Moscow State University, Moscow, Russia}
\affiliation{Institute for High Energy Physics, Protvino, Russia}
\affiliation{Petersburg Nuclear Physics Institute, St. Petersburg, Russia}
\affiliation{Instituci\'{o} Catalana de Recerca i Estudis Avan\c{c}ats (ICREA) and Institut de F\'{i}sica d'Altes Energies (IFAE), Barcelona, Spain}
\affiliation{Stockholm University, Stockholm and Uppsala University, Uppsala, Sweden }
\affiliation{Lancaster University, Lancaster LA1 4YB, United Kingdom}
\affiliation{Imperial College London, London SW7 2AZ, United Kingdom}
\affiliation{The University of Manchester, Manchester M13 9PL, United Kingdom}
\affiliation{University of Arizona, Tucson, Arizona 85721, USA}
\affiliation{University of California Riverside, Riverside, California 92521, USA}
\affiliation{Florida State University, Tallahassee, Florida 32306, USA}
\affiliation{Fermi National Accelerator Laboratory, Batavia, Illinois 60510, USA}
\affiliation{University of Illinois at Chicago, Chicago, Illinois 60607, USA}
\affiliation{Northern Illinois University, DeKalb, Illinois 60115, USA}
\affiliation{Northwestern University, Evanston, Illinois 60208, USA}
\affiliation{Indiana University, Bloomington, Indiana 47405, USA}
\affiliation{Purdue University Calumet, Hammond, Indiana 46323, USA}
\affiliation{University of Notre Dame, Notre Dame, Indiana 46556, USA}
\affiliation{Iowa State University, Ames, Iowa 50011, USA}
\affiliation{University of Kansas, Lawrence, Kansas 66045, USA}
\affiliation{Kansas State University, Manhattan, Kansas 66506, USA}
\affiliation{Louisiana Tech University, Ruston, Louisiana 71272, USA}
\affiliation{Boston University, Boston, Massachusetts 02215, USA}
\affiliation{Northeastern University, Boston, Massachusetts 02115, USA}
\affiliation{University of Michigan, Ann Arbor, Michigan 48109, USA}
\affiliation{Michigan State University, East Lansing, Michigan 48824, USA}
\affiliation{University of Mississippi, University, Mississippi 38677, USA}
\affiliation{University of Nebraska, Lincoln, Nebraska 68588, USA}
\affiliation{Rutgers University, Piscataway, New Jersey 08855, USA}
\affiliation{Princeton University, Princeton, New Jersey 08544, USA}
\affiliation{State University of New York, Buffalo, New York 14260, USA}
\affiliation{Columbia University, New York, New York 10027, USA}
\affiliation{University of Rochester, Rochester, New York 14627, USA}
\affiliation{State University of New York, Stony Brook, New York 11794, USA}
\affiliation{Brookhaven National Laboratory, Upton, New York 11973, USA}
\affiliation{Langston University, Langston, Oklahoma 73050, USA}
\affiliation{University of Oklahoma, Norman, Oklahoma 73019, USA}
\affiliation{Oklahoma State University, Stillwater, Oklahoma 74078, USA}
\affiliation{Brown University, Providence, Rhode Island 02912, USA}
\affiliation{University of Texas, Arlington, Texas 76019, USA}
\affiliation{Southern Methodist University, Dallas, Texas 75275, USA}
\affiliation{Rice University, Houston, Texas 77005, USA}
\affiliation{University of Virginia, Charlottesville, Virginia 22901, USA}
\affiliation{University of Washington, Seattle, Washington 98195, USA}
\author{V.M.~Abazov} \affiliation{Joint Institute for Nuclear Research, Dubna, Russia}
\author{B.~Abbott} \affiliation{University of Oklahoma, Norman, Oklahoma 73019, USA}
\author{B.S.~Acharya} \affiliation{Tata Institute of Fundamental Research, Mumbai, India}
\author{M.~Adams} \affiliation{University of Illinois at Chicago, Chicago, Illinois 60607, USA}
\author{T.~Adams} \affiliation{Florida State University, Tallahassee, Florida 32306, USA}
\author{G.D.~Alexeev} \affiliation{Joint Institute for Nuclear Research, Dubna, Russia}
\author{G.~Alkhazov} \affiliation{Petersburg Nuclear Physics Institute, St. Petersburg, Russia}
\author{A.~Alton$^{a}$} \affiliation{University of Michigan, Ann Arbor, Michigan 48109, USA}
\author{G.~Alverson} \affiliation{Northeastern University, Boston, Massachusetts 02115, USA}
\author{G.A.~Alves} \affiliation{LAFEX, Centro Brasileiro de Pesquisas F{\'\i}sicas, Rio de Janeiro, Brazil}
% ...
%
%
\collaboration{The D0 Collaboration\footnote{with visitors from
$^{a}$Augustana College, Sioux Falls, SD, USA,
$^{b}$The University of Liverpool, Liverpool, UK,
$^{c}$SLAC, Menlo Park, CA, USA,
$^{d}$University College London, London, UK,
$^{e}$Centro de Investigacion en Computacion - IPN, Mexico City, Mexico,
$^{f}$ECFM, Universidad Autonoma de Sinaloa, Culiac\'an, Mexico,
and 
$^{g}$Universit{\"a}t Bern, Bern, Switzerland.%
}} \noaffiliation
\vskip 0.25cm
 
\date{March 23, 2011}

\title{Search for flavor changing neutral currents in decays of top quarks}

\begin{abstract}

We present a search for flavor changing neutral currents 
in decays of top quarks. The analysis is based on a search for
$t\bar{t}\rightarrow\ell'\nu\ell\bar{\ell}$+jets ($\ell, \ell' = e,\mu$) final 
states using 4.1~{\rm fb}$^{-1}$ of integrated luminosity of $p\bar{p}$
collisions at $\sqrt{s} = 1.96$~{\rm TeV}. We extract limits on the branching 
ratio $B(t\rightarrow Zq)$ ($q = u, c$ quarks), assuming anomalous $tuZ$ 
or $tcZ$ couplings. We do not observe any sign of such anomalous coupling and 
set a limit of $B < 3.2\%$ at 95\% C.L.

\end{abstract}

\maketitle

Flavor changing neutral currents (FCNC) allow for transitions between 
quarks of different flavor but same electric charge. FCNC are sensitive 
indicators of physics beyond the standard model (BSM), because 
they are suppressed in the standard model (SM). 
				   
In this paper, we search for FCNC decays of the top ($t$) 
quark~\cite{topdecay}. Within the SM the top quark decays into a 
$W$ boson and a $b$ quark with a rate proportional to the 
Cabibbo-Kobayashi-Maskawa (CKM) matrix element squared, $|V_{\rm tb}|^2$~\cite{topdecay}.
Under the assumption of three fermion families and a unitary $3 \times 3$
CKM matrix, the $|V_{\rm tb}|$ element is severely constrained to  
$|V_{\rm tb}| = 0.999152^{+0.000030}_{-0.000045}$~\cite{pdg}.
While the 
SM branching fraction for $t\rightarrow Zq$ ($q = u, c$ quarks) is predicted 
to be $\approx 10^{-14}$~\cite{tZqfcnc}, supersymmetric 
extensions of the SM with or without $R$-parity violation, or 
quark compositeness predict branching fractions as high as 
$\approx 10^{-4}$~\cite{tZqfcnc,BSMtop1,fcnclhc}. The observation 
of the FCNC decay $t \rightarrow Zq$
would  therefore provide evidence of contributions from BSM physics.

We analyze top-pair production ($t\overline{t}$), where either one or
both of the top quarks decay via $t \rightarrow Zq$ or their charge conjugates 
(hereafter implied). Any top quark that does not decay via $t \rightarrow Zq$ 
is assumed to decay via $t \rightarrow Wb$. We assume that the $t
\rightarrow Zq$ decay is 
generated by an anomalous FCNC term added to the SM
Lagrangian
\begin{eqnarray}
\label{eq:fcnc_lagrangian}
\lefteqn{{\cal L}_{\rm FCNC} =} \nonumber\\ 
& \frac{e}{2 \sin\theta_W \cos\theta_W} \ \bar{t} \, 
   \gamma_\mu ( v_{tqZ} - a_{tqZ} \gamma_5 )\ q \, Z^\mu \,\, + h.c.,
\end{eqnarray}
where $q$, $t$, and $Z$ are the quantum fields for up or charm quarks, top
quarks, and for the $Z$ boson, respectively, $e$ is the electric charge,
and $\theta_W$ the Weinberg angle. We thereby introduce dimension-4
vector, $v_{tqZ}$, and axial vector, $a_{tqZ}$, couplings as defined
in~\cite{fcnc_coupling}. We find in~\cite{nloqcd1, nloqcd2} that the 
next-to-leading order (NLO) effects due to perturbative QCD corrections are negligible
when extracting the branching ratio limits to the leading order (LO)
in Eq.~\ref{eq:fcnc_lagrangian}.

\begin{figure}[tbp!]
\begin{center}
  \includegraphics[width=0.28\textwidth]{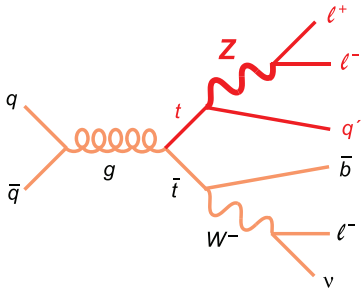}
\end{center}
\caption{Lowest-order diagram for FCNC $t\overline{t} \rightarrow 
         WbZq'$ production, where $q'$ can be either a $u$ or $c$ quark, and 
	 the $W$ and $Z$ bosons decay leptonically.}
\label{fig:FCNCFeyn}
\end{figure}

We investigate channels where the $W$ and $Z$ bosons decay leptonically, 
as shown in Fig.~\ref{fig:FCNCFeyn}. The $u$, $c$, 
and $b$ quarks subsequently hadronize, giving rise to a final state with three charged 
leptons ($\ell = e, \mu$), an imbalance in momentum transverse to the $p\bar{p}$ collision axis 
($\MET$, assumed to be from the escaping neutrino in the $W\rightarrow \ell\nu$ decay), 
and jets. 

This is the first search for FCNC in $t\bar{t}$ decays with trilepton
final states. This mode provides a distinct signature with low background,
albeit at the cost of statistical power. The first measurement 
($b \rightarrow s\gamma$) was published in 1995 by the CLEO Collaboration~\cite{cleo1995}. 
Numerous studies have been done since then to search for FCNC processes in meson 
decays, i.e., $b \rightarrow Zs$ in $B^{+} \rightarrow K^{*+}\ell^{+} 
\ell^{-}$~\cite{Bmeson_cdf, Bmeson_belle, Bmeson_babar}, 
$B \rightarrow K^{*}\nu\bar{\nu}$~\cite{kamenik_arXiv}, and 
$B_{s,d} \rightarrow \ell^{+} \ell^{-}$~\cite{Bmeson_artuso, Bmeson_antonelli}
or $s \rightarrow Zd$ in $K^{+} \rightarrow \pi^{+} \nu \bar{\nu}$~\cite{Kmeson_e949}. 
Using the $D^{+} \rightarrow \pi^{+} \mu^{+} \mu^{-}$ final state in 
$1.3$~{\rm fb}$^{-1}$ of integrated luminosity, the D0 Collaboration has 
set the best branching ratio (B) limits on the FCNC $c\rightarrow Zu$ process 
at $B(D^{+} \rightarrow \pi^{+} \mu^{+} \mu^{-} ) < 3.9 \times 10^{-6}$ at 
90\% C.L.~\cite{fcnc_Dmeson}. 
There are theoretical arguments as to why top quark decays may be the best way to 
study flavor violating couplings of mass-dependent interactions~\cite{rsmodelitself, fcnc_rsmodel}. 
FCNC $tqZ$ and $t q \gamma$ couplings have been studied by the 
CERN $e^+e^-$ Collider (LEP), DESY $ep$ Collider (HERA), and Fermilab $p\bar{p}$ Collider 
(Tevatron) experiments~\cite{fcnc_lep,fcnc_h1,fcnc_zeus,fcnc_tqgamma_cdf,cdflimits}.
The D0 Collaboration has recently published limits on the branching ratios 
determined from FCNC gluon-quark couplings using single top quark final 
states~\cite{d0glimits}. 
The $95\%$ C.L. upper limit on the branching 
ratio of $t \rightarrow Zq$ from the CDF Collaboration uses 
1.9~{\rm fb}$^{-1}$ of integrated luminosity, assumes a top quark mass of $m_{\rm t} = 
175$~{\rm GeV} and uses the measured cross section of $\sigma_{t\bar{t}}
= 8.8 \pm 1.1$ {\rm pb}~\cite{cdflimits}. This result excludes branching ratios of 
$B(t\rightarrow Zq) > 3.7\%$, with an expected limit of 5.0\% $\pm$ 2.2\%. 
To obtain these results, CDF exploited the two lepton plus four jet final
state. This signature occurs when one of the pair-produced 
top quarks decays via FCNC to $Zq$, followed by the decay $Z \rightarrow ee$ 
or $Z \rightarrow \mu\mu$. The other top quark decays to $Wb$, followed 
by the hadronic decay of the $W$ boson. This dilepton signature suffers from large 
background, but profits from more events relative to the trilepton 
final states investigated in this Letter. 

This analysis is based on the measurement of the
$WZ$ production cross section in $\ell\nu\ell\ell$ final states~\cite{WZpub} 
using 4.1~{\rm fb}$^{-1}$ of integrated luminosity of
$p\bar{p}$ collisions at $\sqrt{s} = 1.96$~{\rm TeV}. 
We extend the selection by analyzing events with any number of jets  
in the final state and investigate
observables that are sensitive to the signal topology in order
to select events with $WZ \rightarrow \ell\nu\ell\ell$ decays that  
originate from the pair production of top quarks.

A detailed description of the D0 detector can be found 
elsewhere~\cite{run2det}. Here, we give a brief overview of the main 
sub-systems of the detector. The innermost part is a central tracking 
system surrounded  by a 1.9~{\rm T} superconducting solenoidal magnet. The two 
components of the central tracking system, a silicon microstrip tracker 
and a central fiber tracker, are used to reconstruct $p \bar{p}$ interaction vertices 
and provide the measurement of the momentum of charged particles within 
the pseudorapidity range of $|\eta| < 2$ (with $\eta$ defined relative 
to the center of the detector). The tracking system and magnet are followed 
by the calorimetry that consists of central (CC) and end (EC) 
electromagnetic and hadronic uranium-liquid argon sampling calorimeters, 
and an intercryostat detector (ICD). The CC and two EC calorimeters cover 
$|\eta| < 1.1$ and $1.5 < |\eta| < 4.2$, respectively, while the ICD 
covers $1.1 < |\eta| < 1.4$. The calorimeter measures energies 
of hadrons, electrons, and photons. The muon system consists of a layer of
drift tubes and scintillation counters inside a 1.8~{\rm T} toroidal
magnet. Two similar layers are outside of the toriodal magnet and the
entire system covers $|\eta| < 2$.

An electron is identified from the properties of clusters of 
energy deposited in the CC,
EC, or ICD that match a track reconstructed in the central
tracker. Because of the lack of far forward coverage of the tracker, we 
define EC electrons only within $1.5 < |\eta| < 2.5$.
The calorimeter clusters in the CC and EC are required to pass the isolation cut 
$$\frac{E_{\rm tot} (\Delta {\cal R} < 0.4)  - 
E_{\rm EM} (\Delta {\cal R} < 0.2)}{E_{\rm EM} (\Delta {\cal R} < 0.2)} < 0.1$$
for ``loose" electrons and $< 0.07$ for ``tight" electrons,
where $E_{\rm tot}$ is the total energy in the EM and hadronic calorimeters,  
$E_{\rm EM}$ is the energy found in the EM calorimeter only, and
$\Delta {\cal{R}} = \sqrt{(\Delta\phi)^2+ (\Delta\eta)^2}$, where $\phi$ is 
the azimuthal angle. For the intercryostat region (ICR), $1.1 < |\eta| < 1.5$, 
we form clusters from the energy deposits in the CC, ICD, or EC detectors. These 
clusters are identified as electrons if they pass a neural network requirement 
that is based on the characteristics of the shower and associated track information. 
A muon candidate is reconstructed from track segments within the muon system that 
are matched to a track reconstructed in the central tracker. The trajectory of the 
muon candidate must be isolated from other tracks within a cone of $\Delta {\cal {R}} < 0.5$, 
with the sum of the tracks' transverse momenta, $p_{T}$, in a cone less 
than $4.0~{\rm GeV}$ for ``loose" muons and less than $2.5~{\rm GeV}$ for ``tight" muons.  
Tight muon candidates must also have less than $2.5~{\rm GeV}$ of calorimeter energy in
an annulus of $0.1 < \Delta {\cal {R}} < 0.4$. Jets are reconstructed from 
the energy deposited in the CC and EC calorimeters, using the ``Run II midpoint cone'' 
algorithm~\cite{RunIIcone} of size $\Delta {\cal{R}} = 0.5$, within 
$|\eta| < 2.5$. 

Monte Carlo (MC) samples of $WZ$ and $ZZ$ background events are produced 
using the {\sc pythia} generator~\cite{pythia}. The production of the
$W$ and $Z$ bosons in association with jets ($W + $jets, $Z + $jets), 
collectively referred to as $V + $jets, as well as $t\bar{t}$ processes 
are generated using {\sc alpgen}~\cite{alpgen} interfaced with {\sc pythia} 
for parton evolution and hadronization. In all samples the CTEQ6L1 parton 
distribution function (PDF) set is used, along with $m_{\rm t} = 
172.5$~{\rm GeV}. The $t\bar{t}$ cross section is set to the SM
value at this top quark mass, i.e., $\sigma_{t\bar{t}} = 7.46 ^{+0.48}
_{-0.67}$ {\rm pb}~\cite{ttbar-cross-sec}. This uncertainty is mainly 
due to the scale dependence, PDFs, and the experimental uncertainty on 
$m_{\rm t}$~\cite{top_wa}.

All MC samples are passed through a {\sc geant}~\cite{geant} simulation of 
the D0 detector and overlaid with data events from random beam crossings 
to account for the underlying event. The samples are then corrected for the luminosity 
dependence of the trigger, reconstruction efficiencies in data, and
the beam position. All MC samples are normalized to the luminosity in data 
using NLO calculations of the cross sections, and are subject to the same 
selection criteria as applied to data.

The signal process is generated using the {\sc pythia} generator with
the decay $t\rightarrow Zq$ added. The $Z$ boson helicity is
implemented by reweighting an angular distribution of the
positively charged lepton in the decay $t \to Z q \to \ell^+ \ell^- q$ 
using {\comphep}~\cite{comphep}, modified by the addition of  
the Lagrangian of Eq.~\ref{eq:fcnc_lagrangian}.
The variable $\cos\theta^*$ used for the reweighting is
defined by the angle $\theta^*$ between the $Z$ boson's momentum 
in the top quark rest frame and the momentum of the
positively charged lepton in the $Z$ boson rest frame. 
We assume in the analysis that the vector and axial vector couplings,
as introduced in Eq.~\ref{eq:fcnc_lagrangian}, are
identical to the corresponding couplings for neutral currents (NC) in the SM,
i.e., $v_{tuZ}=1/2-4/3 \sin^2\theta_W = 0.192$ and $a_{tuZ}=1/2$,
where $\sin^2\theta_W=0.231$.
To study the influence of different values of the couplings, we
also analyse the following cases: (i.a) $v_{tuZ}=1, a_{tuZ}=0$; 
(i.b) $v_{tuZ}=0, a_{tuZ}=1$; and (ii) $v_{tuZ}=a_{tuZ}=1/\sqrt{2}$. 
As expected, the first two give identical results. 
The difference obtained by using cases (i), (ii), and using the 
values of the SM NC couplings is included as
systematic uncertainty. Therefore, our
result is independent of the actual values of $v_{tqZ}$ and $a_{tqZ}$.
Since we do not distinguish $c$ and $u$ quark jets
our results are valid also for $u$ and $c$ quarks separately.

The total selection efficiency, calculated as a function of 
$B =  \Gamma(t\rightarrow Zq)/\Gamma_{\rm tot}$, where $\Gamma_{\rm tot}$ contains 
$t\rightarrow Wb$ and $t\rightarrow Zq$ decays only, can be written as
\begin{eqnarray}
\label{eq:effttbar}
\epsilon_{t\bar{t}} &=& (1-B)^2 \cdot \epsilon_{t\bar{t} \to W^{+} b W^{-}
  \bar{b}} \\ \nonumber
&+& 2B(1-B) \cdot \epsilon_{t\bar{t} \to Z q W^{-} \bar{b}}
+ B^2 \cdot \epsilon_{t\bar{t} \to Z q Z \bar{q}},
\end{eqnarray}
where the efficiency $\epsilon_{t\bar{t} \to W^{+} b W^{-} \bar{b}}$
for the SM $t\bar{t}$ background contribution is used, 
along with the efficiencies $\epsilon_{t\bar{t} \to Z q W^- \bar{b}}$ 
and $\epsilon_{t\bar{t} \to Z q Z \bar{q}}$
that include the FCNC top quark decays.

We consider four independent decay signatures: $eee+\MET + X$,
$ee\mu+\MET + X$, $\mu\mu e+\MET + X$, and $\mu\mu\mu+\MET + X$,
where $X$ is any number of jets. We require the events 
to have at least three lepton candidates with $p_T > 15$~{\rm GeV} that 
originate from the same $p\bar{p}$ interaction vertex and are separated from 
each other by $\Delta {\cal{R}} > 0.5$. Jets are excluded from consideration 
unless they have $p_T>20$~{\rm GeV}. 
We also require that the jets are separated from electrons
by $\Delta {\cal{R}} > 0.5$. There is no fixed separation cut between the 
muon and jets but the muon 
isolation requirement rejects most muons within $\Delta {\cal{R}} < 0.4$ of a jet. 
The event must also have 
$\MET > 20$~{\rm GeV}, which is calculated from the energy found in the calorimeter cells 
and $p_{T}$ corrected for any muons reconstucted in the event. Furthermore, all energy 
corrections applied to electrons and jets are propagated through to the $\MET$. 

Events are selected using triggers based on electrons and muons. There 
are several high-$p_T$ leptons from the decay of the heavy gauge 
bosons providing a total trigger efficiency for all signatures of 
$98\% \pm 2\%$.

To identify the leptons from the $Z$ boson decay, we consider
only pairs of electrons or muons, additionally requiring them to 
have opposite electric charges. 
If no lepton pair is found within the invariant mass intervals of 74--108~{\rm GeV} 
($ee$), 65--115~{\rm GeV} ($\mu\mu$) or 60--120~{\rm GeV} ($ee$, with one electron 
in the ICR) the event is rejected, else, the pair that has an invariant mass closest to 
the $Z$ boson mass $M_{Z}$ is selected as the $Z$ boson. The lepton with the highest 
$p_{T}$ of the remaining muons or CC/EC electrons in the event is selected 
as originating from the $W$ boson decay. From simulation, this assignment of the 
three charged leptons to $Z$ and $W$ bosons is found to be  
$\approx$ 100\% correct for $ee\mu$ and $\mu\mu e$, and about 
92\% and 89\% for the $eee$ and $\mu\mu\mu$ channels, respectively.

Thresholds in the selection criteria are the same as in Ref.~\cite{WZpub} 
and the acceptance multiplied by efficiency results are summarized in 
Table~\ref{Axeff} for the FCNC signal. These values are calculated with respect 
to the total rate expected for all three generations of leptonic $W$ and $Z$ decays.

\begin{table*}[t]
\begin{tabular}{|l|c|c|c|c|} \hline \hline

$n_{\rm jet}$ & Inclusive       & $0$                           & $1$             & $\geq 2$        \\ \hline 
Channel       & \multicolumn{4}{c|}{$\epsilon_{t\bar{t} \to ZqW^-\bar{b}}$ (\%)} \\ \hline
$eee$         & $1.65 \pm 0.24$ & $(7.65 \pm 1.45)\cdot10^{-2}$ & $0.57 \pm 0.09$ & $1.00 \pm 0.15$ \\ \hline
$ee\mu$       & $1.92 \pm 0.18$ & $(6.77 \pm 1.05)\cdot10^{-2}$ & $0.58 \pm 0.06$ & $1.17 \pm 0.11$ \\ \hline
$\mu \mu e$   & $1.23 \pm 0.13$ & $(3.37 \pm 0.73)\cdot10^{-2}$ & $0.34 \pm 0.04$ & $0.84 \pm 0.10$ \\ \hline
$\mu \mu \mu$ & $1.48 \pm 0.19$ & $(3.05 \pm 0.74)\cdot10^{-2}$ & $0.38 \pm 0.06$ & $1.06 \pm 0.15$ \\ \hline
Channel       & \multicolumn{4}{c|}{$\epsilon_{t\bar{t} \to ZqZ\bar{q}}$ (\%)} \\ \hline
$eee$         & $1.22 \pm 0.18$ & $(4.69 \pm 0.68)\cdot10^{-2}$ & $0.41 \pm 0.06$ & $0.76 \pm 0.11$ \\ \hline
$ee\mu$       & $3.75 \pm 0.38$ & $(1.07 \pm 0.11)\cdot10^{-1}$ & $1.08 \pm 0.11$ & $2.56 \pm 0.25$ \\ \hline
$\mu \mu e$   & $1.47 \pm 0.16$ & $(3.22 \pm 0.57)\cdot10^{-2}$ & $0.38 \pm 0.05$ & $1.06 \pm 0.32$ \\ \hline
$\mu \mu \mu$ & $2.76 \pm 0.36$ & $(3.63 \pm 0.69)\cdot10^{-2}$ & $0.63 \pm 0.09$ & $2.10 \pm 0.28$ \\ \hline \hline
\end{tabular}
\caption{Final efficiencies in \% including detector and kinematic acceptance as
 well as detector efficiencies for each decay signature as a function of jet 
 multiplicity $n_{\rm jet}$. The efficiency, $\epsilon$, is defined assuming fully 
 leptonic decays of the vector bosons from top quarks, 
 as defined as in Eq.~\ref{eq:effttbar}.  
 The statistical and systematic uncertainties have been added in quadradure.}
\label{Axeff}
\end{table*}

In addition to SM $WZ$ production, the other major background is from processes 
with a $Z$ boson and an additional object misidentified as the lepton 
from the $W$ boson decay (e.g., from $Z + $jets, $ZZ$, and $Z \gamma$). 
A small background contribution is expected from processes such as 
$W + $jets and SM $t\bar{t}$ production.
The $WZ$, $ZZ$, and $t\bar{t}$ backgrounds are estimated from the 
simulation, while the $V + $jets and $Z\gamma$ backgrounds are estimated 
using data-driven methods.

One or more jets in $V + $jets events can be misidentified as a lepton from 
$W$ or $Z$ boson decays. To estimate this contribution, 
we define a {\it false} lepton category for electrons and muons. 
A {\it false} electron is required to have most of its energy deposited 
in the electromagnetic part of the calorimeter and satisfy 
calorimeter isolation criteria for electrons, but have a shower shape 
inconsistent with that of an electron. A muon candidate is categorized 
as {\it false} if it fails isolation criteria, as determined from the 
total $p_{T}$ of tracks located within a cone $\Delta {\cal{R}} = 0.5$ 
around the muon. These requirements ensure that the {\it false} lepton 
is either a misidentified jet or a lepton from the semi-leptonic decay 
of a heavy-flavor quark. Using a sample of data events, collected using 
jet triggers with no lepton requirement, 
we measure the ratio of misidentified leptons passing two different 
selection criteria, {\it false} lepton and signal lepton, 
as a function of $p_T$ in three bins, $n_{\rm jet}$ = 0, 1, 
and $\geq 2$, where $n_{\rm jet}$ is the number of jets. 
We then select a sample of $Z$ boson decays with at least one additional {\it false} 
lepton candidate for each final state signature. The contribution from 
the $V + $jets background is estimated by scaling the number of events 
in this $Z+${\it false} lepton sample by the corresponding $p_T$-dependent 
misidentification ratio.

Initial or final state radiation in $Z\gamma$ events can mimic the 
signal process if the photon either converts into an $e^{+}e^{-}$ pair 
or is wrongly matched with a central track mimicking an electron and the 
$\MET$ is mismeasured. As a result the $Z\gamma$ process is a background to the  
final state signatures with $W\to e\nu$ decays. To estimate the contribution 
from this background, we model the kinematics of these events using the $Z\gamma$ 
NLO MC simulation~\cite{Baur}.  We scale this result by the rate at which a photon is 
misidentified as an electron. This rate is obtained using a data sample 
of $Z\to\mu\mu$ events containing a radiated photon, as it offers an almost 
background-free source of photons. The invariant mass $M(\mu\mu\gamma)$ is
reconstructed and required to be consistent with the $Z$ boson mass. 
The $Z\rightarrow \mu\mu$ decay is chosen to avoid any ambiguity when 
assigning the electromagnetic shower to the final state photon candidate. 
As the $Z\gamma$ NLO MC does not model recoil jets, {\sc pythia} MC samples are used to 
estimate $Z\gamma$ background jet multiplicities and $\MET$.  As the {\sc pythia} 
samples do not contain events with final state radiation, we find the fraction of 
$Z \gamma$ events in data and {\sc pythia} MC that pass our $\MET$ cut and 
take the difference as a systematic uncertainty.

After all selection criteria have been applied, we observe a total of $35$ candidate events and 
expect $31.7 \pm 0.3({\rm stat}) \pm 3.9 ({\rm syst})$ background events from SM processes. 
The statistical uncertainty is due to MC statistics while the sources of systematic 
uncertainties are discussed later.
Table~\ref{tab:EvYieldsJetBins} summarizes the number of events in each $n_{\rm jet}$
bin. The observed number of candidate and background events for each topology, summing
over $n_{\rm jet}$, are summarized in Table~\ref{tab:EvYieldsInclJets}. 
In Tables~\ref{tab:EvYieldsJetBins} and~\ref{tab:EvYieldsInclJets} 
and in all the following figures, we assume a $B$ of 5\%.

\begin{table*}[t]
\begin{tabular}{|c|c|c|c|} \hline \hline

$n_{\rm jet}$   & $0$                       & $1$                      & $\geq 2$                  \\ \hline
Background      & $25.66 \pm 0.28 \pm 3.26$ & $5.06 \pm 0.14 \pm 0.56$ & $0.92 \pm 0.08 \pm 0.09$  \\ \hline
$t\bar{t}\rightarrow WbZq$ & $0.20 \pm 0.03$   & $1.80 \pm 0.27$   & $3.87 \pm 0.56$ \\ 
$t\bar{t}\rightarrow ZqZq$ & $0.002 \pm 0.001$ & $0.020 \pm 0.003$ & $0.050 \pm 0.007$ \\ \hline
Observed        & 30                        & 4                        & 1                         \\ \hline \hline
\end{tabular}
\caption{Number of observed events, expected number of $t\bar{t}$ FCNC events, 
        and number of expected background events for each $n_{\rm jet}$ bin
        with statistical and systematic uncertainties. The MC statistical uncertainty
	on the $t\bar{t}$ signal is negligible, and we only present the systematic uncertainties. 
	We assume $B = 5\%$.}
\label{tab:EvYieldsJetBins}
\end{table*}

\begin{table*}[t]
\begin{tabular}{|l|c|c|c|c|} \hline \hline

Source      & $eee$                    & $ee\mu$                  & $e\mu\mu$                & $\mu\mu\mu$               \\ \hline
$WZ$        & $6.64 \pm 0.07 \pm 1.19$ & $7.51 \pm 0.08 \pm 1.11$ & $4.75 \pm 0.06 \pm 0.69$ & $6.10 \pm 0.07 \pm 1.00$ \\ 
$ZZ$        & $0.33 \pm 0.03 \pm 0.06$ & $1.76 \pm 0.07 \pm 0.17$ & $0.46 \pm 0.04 \pm 0.07$ & $1.30 \pm 0.06 \pm 0.21$ \\ 
$V$ + jets  & $0.60 \pm 0.13 \pm 0.11$ & $0.40 \pm 0.18 \pm 0.17$ & $0.48 \pm 0.10 \pm 0.01$ & $0.22 \pm 0.05 \pm 0.03$ \\ 
$Z\gamma$   & $0.18 \pm 0.05 \pm 0.08$ & $< 0.001$                & $0.66 \pm 0.07 \pm 0.38$ & $< 0.001$        \\ 
$t\bar{t}\rightarrow WbWb$  & $0.04 \pm 0.01 \pm 0.01$ & $0.04 \pm 0.01 \pm 0.01$ & $0.05 \pm 0.01 \pm 0.01$ & $0.04 \pm 0.01 \pm 0.01$ \\ \hline
Background  & $7.89 \pm 0.16 \pm 1.20$ & $9.71 \pm 0.21 \pm 1.14$ & $6.40 \pm 0.14 \pm 0.79$ & $7.66 \pm 0.11 \pm 1.02$ \\ \hline
$t\bar{t}\rightarrow WbZq$  & $1.57 \pm 0.22$   & $1.73 \pm 0.17$   & $1.17 \pm 0.13$   & $1.41 \pm 0.18$\\ 
$t\bar{t}\rightarrow ZqZq$  & $0.010 \pm 0.001$ & $0.029 \pm 0.003$ & $0.011 \pm 0.001$ & $0.022 \pm 0.003$\\ \hline
Observed    & 8                        & 13                       & 9                        & 5        \\ \hline \hline
\end{tabular}
\caption{Number of observed events, expected number of $t\bar{t}$ FCNC events,
        and number of expected background events for each final state signature 
        with statistical and systematic uncertainties. The MC statistical uncertainty
	on the $t\bar{t}$ signal is negligible, and we only present the systematic 
	uncertainties. We assume $B = 5\%$.}
\label{tab:EvYieldsInclJets}
\end{table*}

To achieve better separation between signal and background, we analyze 
the $n_{\rm jet}$ and $H_{\rm T}$ distributions (defined as the scalar sum 
of transverse momenta of all leptons, jets, and $\MET$), and the 
reconstructed invariant mass for the products of the decay $t \rightarrow Zq$.

\begin{figure}[htbp]
 \begin{center}
        {\includegraphics[width=0.45\textwidth]{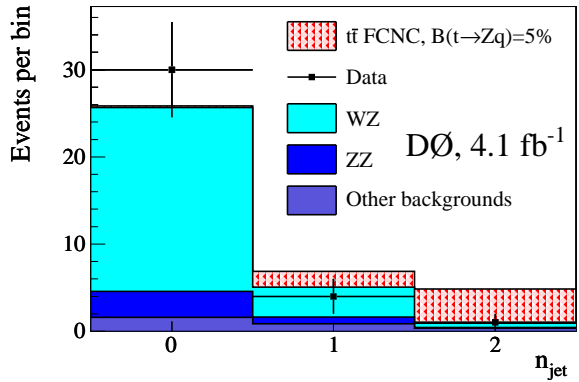}}
	\caption{Distribution of $n_{\rm jet}$ for data, for simulated
	         FCNC $t\bar{t}$ signal, and for the expected 
		 background. The $ZqZq$ signal is included in the $t\bar{t}$ FCNC contribution 
		 but is expected to be small, as can be seen from Tables~\ref{tab:EvYieldsJetBins} 
                 and~\ref{tab:EvYieldsInclJets}.
}
    \label{fig:jetbins}
  \end{center}
\end{figure}

\begin{figure}[htbp]
 \begin{center} 
   \includegraphics[width=0.45\textwidth]{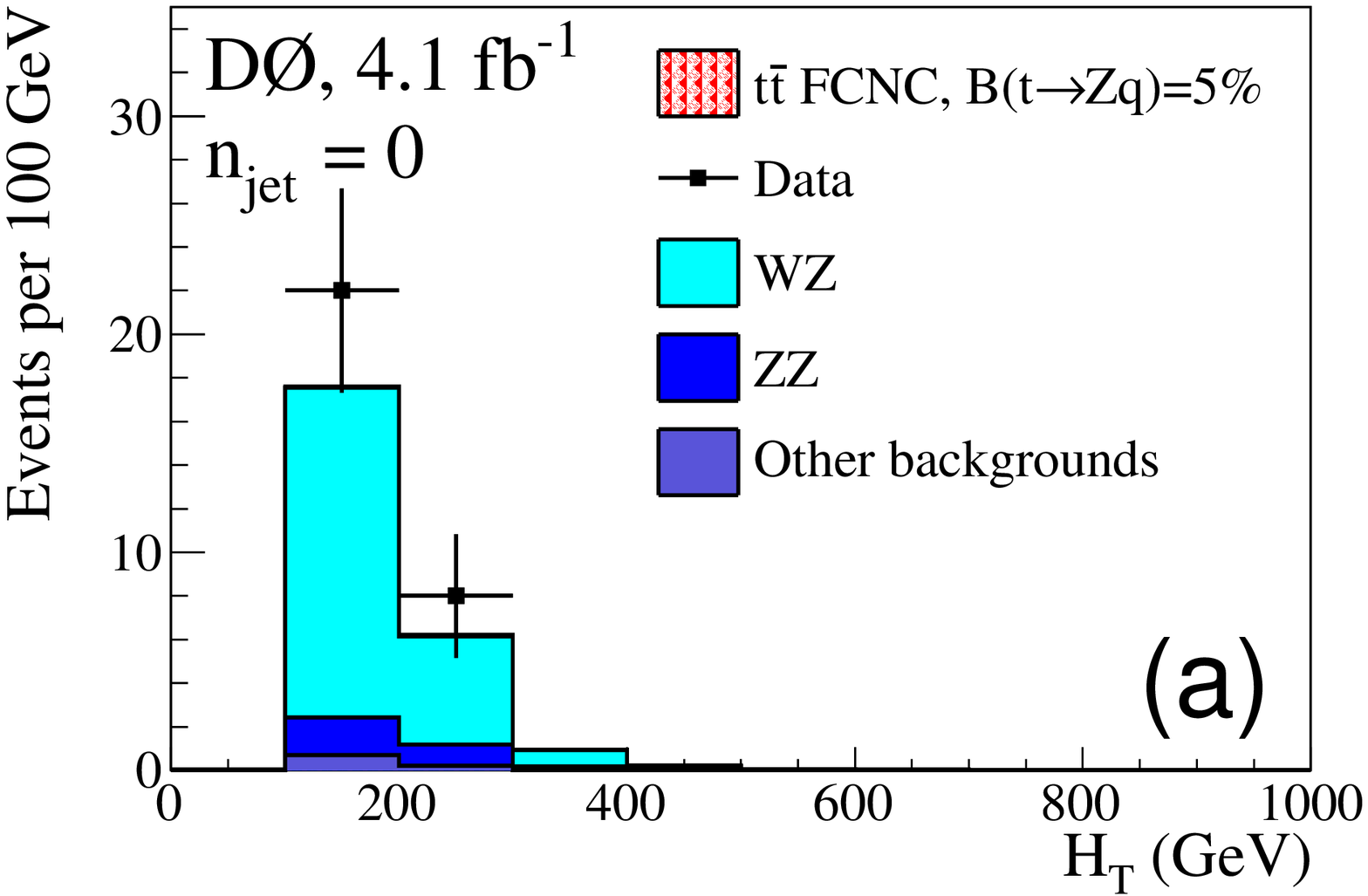}
   \includegraphics[width=0.45\textwidth]{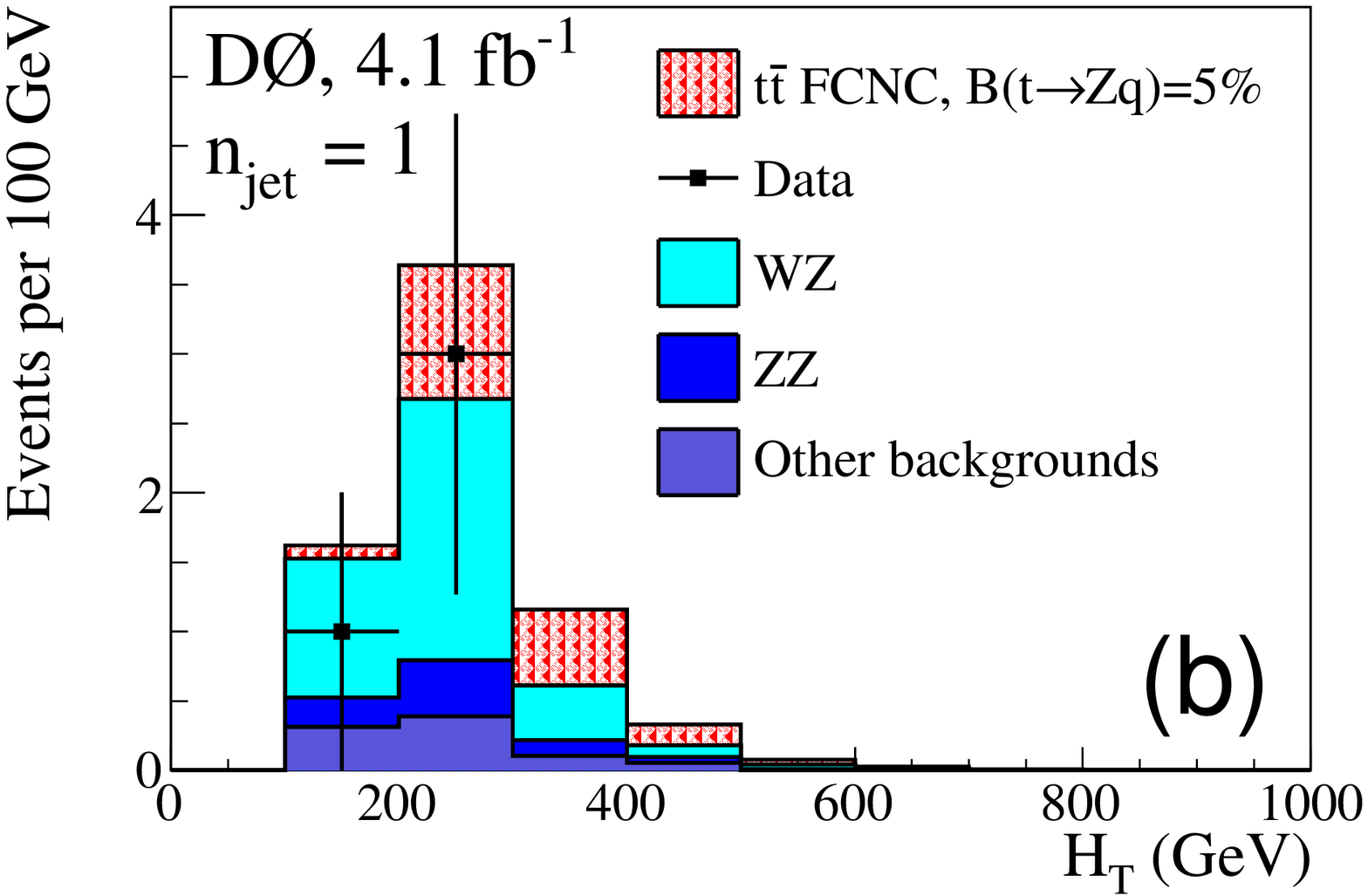}
   \includegraphics[width=0.45\textwidth]{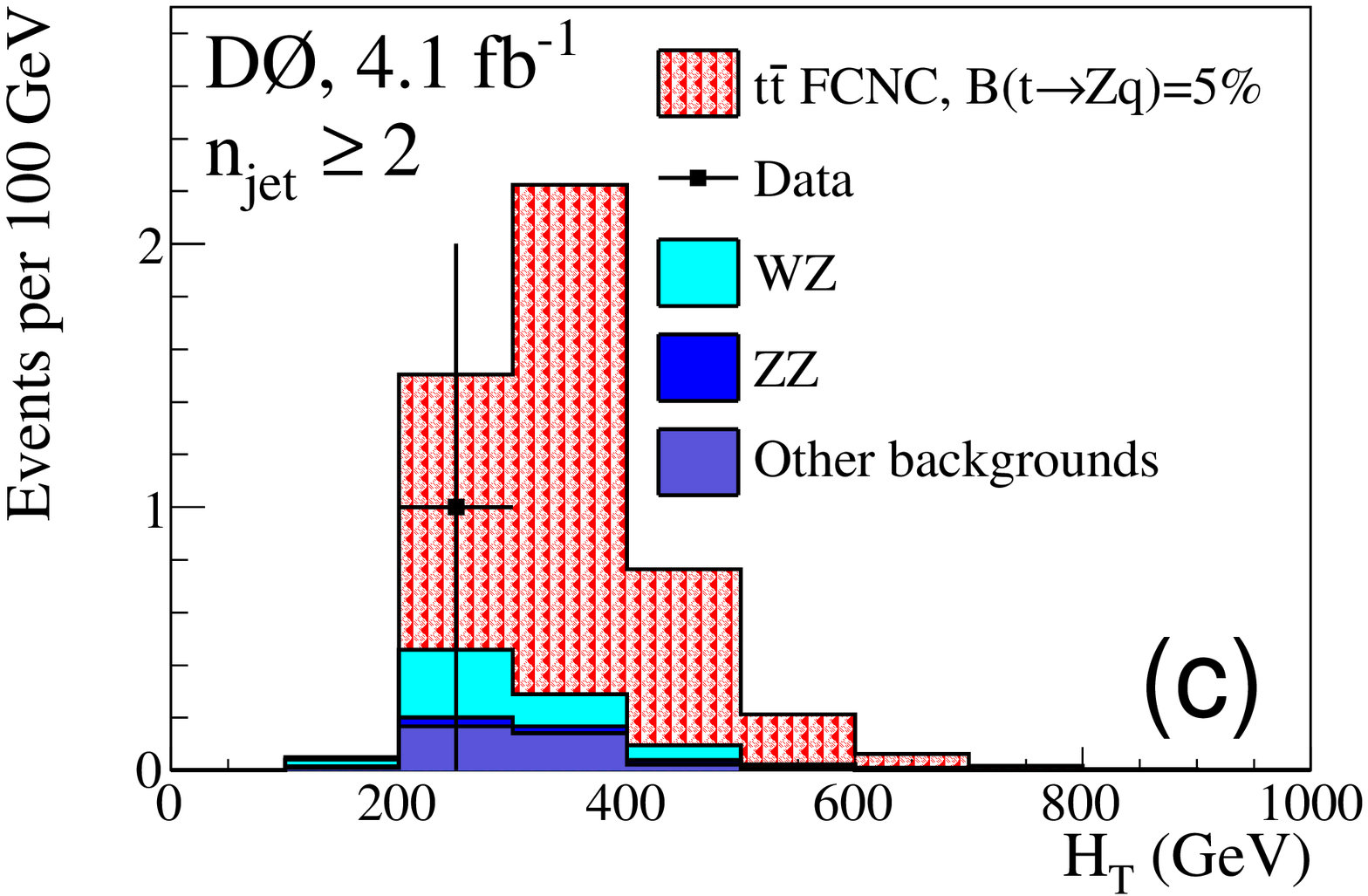}
   \caption{$H_{\rm T}$ distribution of data, FCNC $t\bar{t}$ signal, 
            and expected background for events with (a) $n_{\rm jet} = 0$, 
	    (b) $n_{\rm jet} = 1$, and (c) $n_{\rm jet} \geq 2$. 
}
   \label{fig:ht_all} 
  \end{center} 
\end{figure}

The jet multiplicities in data, SM background, and in FCNC top quark decays 
are shown in Fig.~\ref{fig:jetbins}.  
FCNC $t\bar{t}$ production leads to larger jet 
multiplicities and also a larger $H_{\rm T}$. This is shown in Fig.~\ref{fig:ht_all}.

\begin{figure}[ht]
 \begin{center} 
   \includegraphics[width=0.45\textwidth]{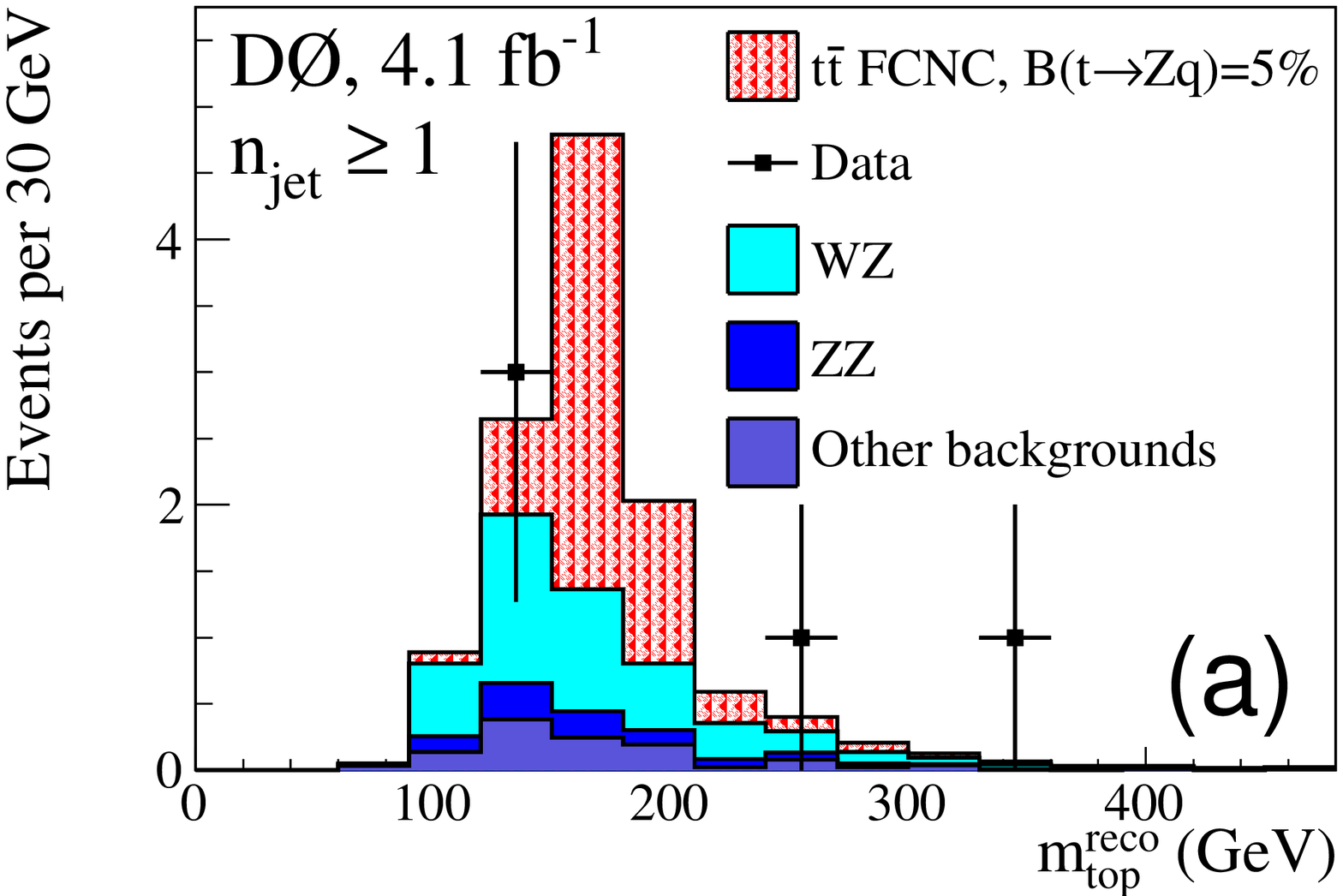}
   \includegraphics[width=0.45\textwidth]{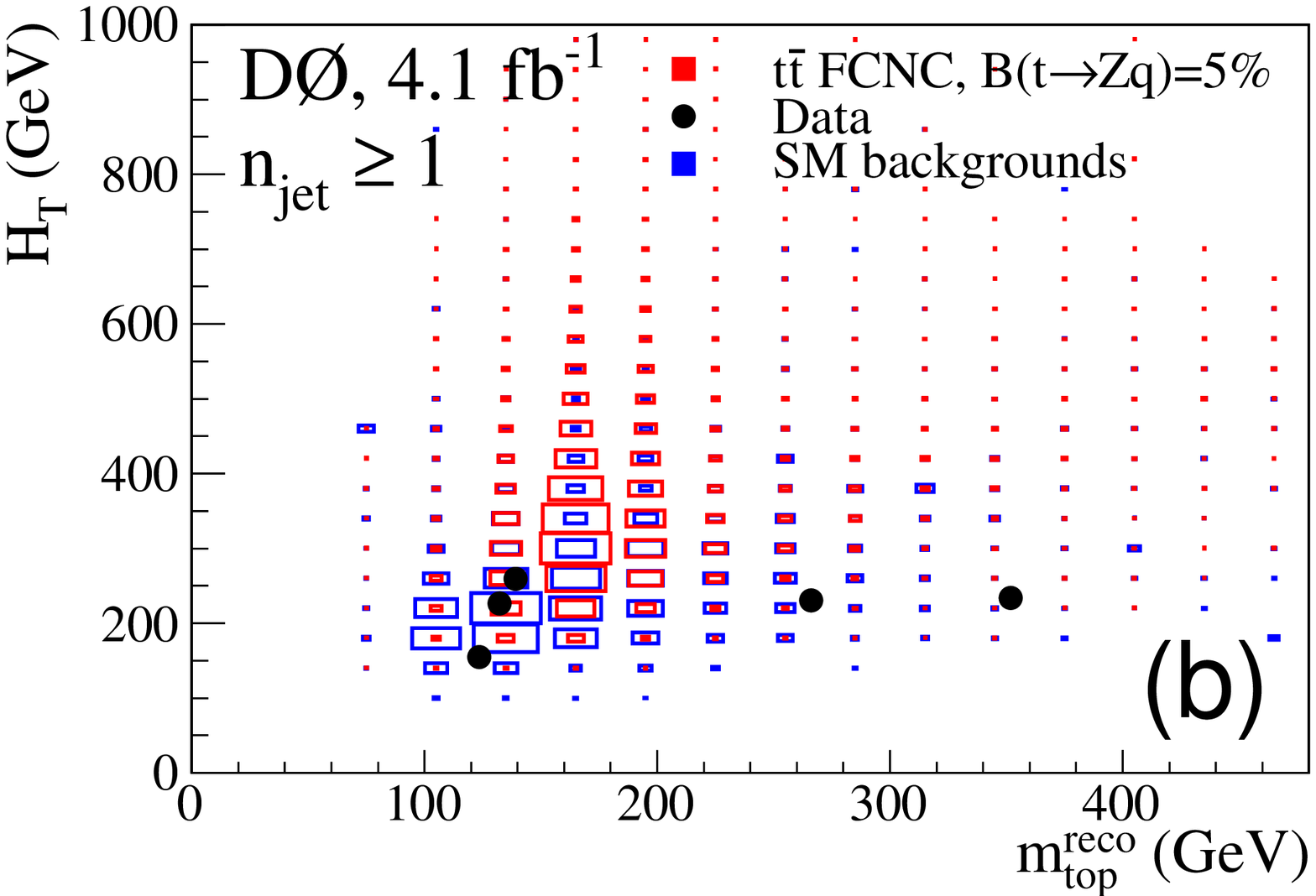}
   \caption{(a) $m_{\rm t}^{\rm reco}$ distribution of data, FCNC $t\bar{t}$ signal, 
            and expected background for events with $n_{\rm jet} \geq 1$;
            (b) $m_{\rm t}^{\rm reco}$ vs. $H_{\rm T}$ distribution of
	    data, FCNC $t\bar{t}$ signal, and background for events with 
	    $n_{\rm jet} \geq 1$. 
}
   \label{fig:mzjet} 
  \end{center} 
\end{figure}

To further increase our sensitivity we reconstruct the mass of the top quark that
decays via FCNC to a $Z$ boson and a quark ($t
\rightarrow Zq$). In events with $n_{\rm jet} = 0$, this
variable is not defined. In events with one jet, we calculate 
the invariant mass, $m_{\rm t}^{\rm reco} \equiv M(Z,{\rm jet})$, from the 4-momenta 
of the jet and the identified $Z$ boson, to reconstruct $m_{\rm t}$. 
For events with two or more jets, we use the jet that gives a 
$m_{\rm t}^{\rm reco}$ closest to $m_t=172.5$~{\rm GeV}. 
The $m_{\rm t}^{\rm reco}$ distribution is shown in Fig.~\ref{fig:mzjet}(a). In 
Fig.~\ref{fig:mzjet}(b), we present a 2-dimensional distribution
of $m_{\rm t}^{\rm reco}$ and $H_{\rm T}$.

None of the observables in 
Figs.~\ref{fig:jetbins} -- \ref{fig:mzjet} show evidence for the presence of FCNC  
in the decay of $t\bar{t}$. We therefore set 95\% C.L. limits on the 
branching ratio $B(t\rightarrow Zq)$. The limits are derived from 10 bins of the
$H_{\rm T}$ distributions for $n_{\rm jet} = 0, 1$, and $\geq 2$. For the
channels with $n_{\rm jet} = 1$ and $n_{\rm jet} \geq 2$, we split each $H_{\rm T}$
distribution into 4 bins in $m_{\rm t}^{\rm reco}$, $m_{\rm t}^{\rm reco} <$ 120~{\rm GeV}, 
120 $< m_{\rm t}^{\rm reco} <$ 150~{\rm GeV}, 150 $< m_{\rm t}^{\rm reco} <$ 
200~{\rm GeV}, and $m_{\rm t}^{\rm reco} >$ 200~{\rm GeV}. 

When calculating the limit on the branching ratio we consider several sources 
of systematic uncertainty.
The systematic uncertainties for lepton-identification efficiencies are 
$15\%$ ($eee$), $11\%$ ($ee\mu$), $9\%$ ($\mu\mu e$), and $12\%$ ($\mu\mu\mu$).
The systematic uncertainty assigned to the choice of PDF is $5\%$.
In addition, we assign $9\%$ systematic uncertainty on 
$\sigma_{t\bar{t}}$~\cite{ttbar-cross-sec}. This includes the dependence on
the uncertainty of $m_{\rm t}$~\cite{top_wa}. Furthermore, $m_{\rm t}$ is changed from 
$172.5$~{\rm GeV} to $175$~{\rm GeV} in $t\bar{t}$ MC samples with the difference 
in the result taken as a systematic uncertainty. 
We vary the $v_{tqZ}$ and $a_{tqZ}$ couplings as explained before Eq.~\ref{eq:effttbar}, 
resulting in a 1\% systematic uncertainty on the acceptance.  
Due to the uncertainty on the theoretical cross sections for $WZ$ and
$ZZ$ production, we assign a $10\%$~\cite{vv-cross-sec} systematic uncertainty to each.
The major sources of systematic uncertainty on the estimated $V + $jets
contribution arise from the \MET~requirement and the statistics in the multijet 
sample used to measure the lepton-misidentification rates. These effects 
are estimated independently for each signature and found to be between $20\%$ and $30\%$. 
The systematic uncertainty on the $Z\gamma$ background is estimated
to be $40\%$ and $58\%$ for the $eee$ and $\mu\mu e$ channels, respectively. 
Uncertainties on jet energy scale, jet energy resolution, jet 
reconstruction, and identification efficiency are estimated by varying 
parameters within their experimental uncertainties. For $n_{\rm jet} = 0$ the 
uncertainty is found to be $1\%$, for $n_{\rm jet} = 1$ it is $5\%$, 
and for $n_{\rm jet} \geq 2$ it is $20\%$. The measured integrated luminosity 
has an uncertainty of $6.1\%$~\cite{lumi}.

We use a  modified frequentist approach~\cite{cls} where the signal 
confidence level $CL_s$, defined as the ratio of the
confidence level for the signal+background hypothesis to the
background-only hypothesis ($CL_s = CL_{s+b}/CL_b$), is calculated by
integrating the distributions of a test statistic over the outcomes
of pseudo-experiments generated according to Poisson statistics for
the signal+background and background-only hypotheses. The test
statistic is calculated as a joint log-likelihood ratio (LLR)
obtained by summing LLR values over the bins of the $H_{\rm T}$ distributions.
Systematic uncertainties are incorporated via Gaussian smearing of
Poisson probabilities for signal and backgrounds in the pseudo-experiments. 
All correlations between signal and backgrounds are maintained. To reduce 
the impact of systematic uncertainties on the sensitivity of the analysis, 
the individual signal and background contributions are fitted to the data,
by allowing a variation of the background (or signal+background) prediction,
within its systematic uncertainties~\cite{collie}. 
The likelihood is constructed 
via a joint Poisson probability over the number of bins in the calculation,
and is a function of scaling factors for the systematic uncertainties, which 
are given as Gaussian constraints associated with their priors.

\begin{figure}[tbp!]
\begin{center}
  \includegraphics[width=0.5\textwidth]{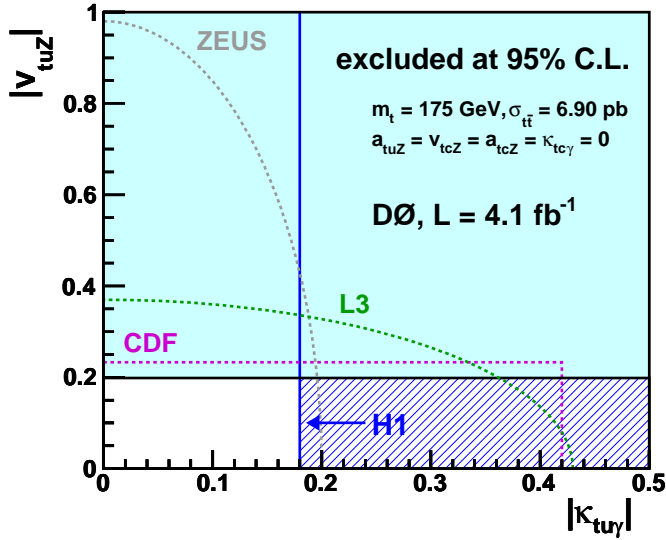}
\end{center}
\caption{Upper limits at the 95\%~C.L. on the anomalous $\kappa_{t u
\gamma}$ and $v_{tuZ}$ couplings assuming
$m_t = 175$~{\rm GeV}. Both D0 and CDF limits on $v_{tqZ}$ are scaled
to the SM cross section of $\sigma_{t\bar{t}} = 6.90$~pb~\cite{ttbar-cross-sec}.
Anomalous axial vector couplings and 
couplings of the charm quark are neglected: $a_{tuZ}  =
v_{tcZ} = a_{tcZ} = \kappa_{t c \gamma} = 0$. The scale parameter for
the anomalous dimension-5 coupling $\kappa_{t u \gamma}$ is set to
$\Lambda = m_t = 175$~{\rm GeV}~\cite{fcnc_h1}. Any dependence of the
Tevatron limits on $\kappa_{t u \gamma}$ is not displayed as the 
change is small and at most 6\% for
$\kappa_{t u \gamma}=0.5$. The domain excluded by D0 is represented by 
the light (blue) shaded area. The hatched area corresponds to the additional domain
excluded at HERA by the H1 experiment~\cite{fcnc_h1}. Also shown are upper limits
obtained at LEP by the L3 experiment~\cite{fcnc_lep} (green dashed), at HERA by the
ZEUS experiment~\cite{fcnc_zeus} (grey dashed), and at the Tevatron by the CDF
experiment~\cite{fcnc_tqgamma_cdf,cdflimits} (magenta dashed). The region above 
or to the right of the respective lines is excluded.}.
\label{fig:colliders}
\end{figure}

We determine an observed limit of $B(t \rightarrow Zq) < 3.2\%$, 
with an expected limit of $<3.8\%$ at the 95\% C.L.
The limits on the branching ratio are converted to limits at the 95\%~C.L. 
on the FCNC vector, $v_{tqZ}$, and axial vector, $a_{tqZ}$,
couplings as defined in Eq.~\ref{eq:fcnc_lagrangian} using the
relation given in~\cite{fcnc_coupling}. This can be done for any
point in the ($v_{tqZ}$, $a_{tqZ}$) parameter space and for different
quark flavors ($u,c$) since the differences
in the helicity structure of the couplings are covered as systematic
uncertainties in the limit on the branching ratio.
Assuming only one non-vanishing $v_{tqZ}$ coupling ($a_{tqZ} = 0$), we 
derive an observed (expected) limit of $v_{tqZ}<0.19$ ($<0.21$) for 
$m_t=172.5$~{\rm GeV}. Likewise, this limit holds assuming
only one non-vanishing $a_{tqZ}$ coupling.
Figure~\ref{fig:colliders} shows current limits from experiments at the
LEP, HERA, and Tevatron colliders as a function of the FCNC couplings
$\kappa_{t u \gamma}$ (defined in Ref.~\cite{fcnc_coupling}) and
$v_{tuZ}$ for $m_t=175$~{\rm GeV}.

In summary, we have presented a search for top quark
decays via FCNC in \ttbar\ events leading to final states
involving three leptons, an imbalance in transverse momentum, and jets.
These final states
have been explored for the first time in the context of FCNC couplings.
In the absence of signal, we expect a limit of $B(t \rightarrow Zq) <
3.8\%$ and set a limit of $B(t \rightarrow Zq) < 3.2\%$ at the $95\%$~C.L. 
which is currently the world's best limit. This translates into
an observed limit on the FCNC coupling of $v_{tqZ}<0.19$ for $m_t=172.5$~{\rm GeV}. 

%
We thank the staffs at Fermilab and collaborating institutions,
and acknowledge support from the
DOE and NSF (USA);
CEA and CNRS/IN2P3 (France);
FASI, Rosatom and RFBR (Russia);
CNPq, FAPERJ, FAPESP and FUNDUNESP (Brazil);
DAE and DST (India);
Colciencias (Colombia);
CONACyT (Mexico);
KRF and KOSEF (Korea);
CONICET and UBACyT (Argentina);
FOM (The Netherlands);
STFC and the Royal Society (United Kingdom);
MSMT and GACR (Czech Republic);
CRC Program and NSERC (Canada);
BMBF and DFG (Germany);
SFI (Ireland);
The Swedish Research Council (Sweden);
and
CAS and CNSF (China).

\end{document}